\documentclass[runningheads]{llncs}
\usepackage{hyperref}
\usepackage{graphicx}
\begin{document}
\title{Motifs, Phrases, and Beyond: The Modelling of Structure in Symbolic Music Generation}
\titlerunning{Motifs, Phrases, and Beyond}
%
\author{Keshav Bhandari\inst{1}\orcidID{0009-0008-2045-6319} \and
Simon Colton\inst{2}\orcidID{0000-0002-4887-6947}}
%
\authorrunning{K. Bhandari and S.Colton}
%
\institute{
\email{k.bhandari@qmul.ac.uk}\\
\and
\email{s.colton@qmul.ac.uk}\\
School of Electronic Engineering and Computer Science\\
Queen Mary University of London
}
\maketitle              
\begin{abstract}
Modelling musical structure is vital yet challenging for artificial intelligence systems that generate symbolic music compositions. This literature review dissects the evolution of techniques for incorporating coherent structure, from symbolic approaches to foundational and transformative deep learning methods that harness the power of computation and data across a wide variety of training paradigms. In the later stages, we review an emerging technique which we refer to as “sub-task decomposition" that involves decomposing music generation into separate high-level structural planning and content creation stages. Such systems incorporate some form of musical knowledge or neuro-symbolic methods by extracting melodic skeletons or structural templates to guide the generation. Progress is evident in capturing motifs and repetitions across all three eras reviewed, yet modelling the nuanced development of themes across extended compositions in the style of human composers remains difficult. We outline several key future directions to realize the synergistic benefits of combining approaches from all eras examined.

\end{abstract}
\section{Introduction}
Musical structure embodies the thoughtful organization of fundamental elements such as pitch, harmony, rhythm, and timbre in a composition \cite{ong_structural_nodate}. Structure gives a sense of coherence, unity, and direction to the piece. On a basic level, notes are organized into motives and phrases. These phrases are then combined into higher-level sections such as verses, choruses and melodic segments. The repetition, variation, and development of musical ideas and themes manifest in the music and create relationships between these sections to form an overarching structure. Understanding and perceiving this musical structure is key to appreciating and enjoying music for listeners \cite{gabrielsson2014relationship}.

Musical structure is undoubtedly a crucial aspect of people's cognitive process that influences their perception of music \cite{lazzari_pitchclass2vec_2023,stevens_music_2012,tan_harmonic_1981}. Surprisingly, research involving infants as young as four and a half years old suggests that musical experience might not be a prerequisite for perceiving musical phrase structures \cite{krumhansl_infants_1990}. Studies focusing on rhythm have delved into perceptual elements like timing and tempo \cite{honing_structure_2002} and syncopated patterns \cite{fitch_perception_2007}. For human listeners, perceiving musical structure can be thought of as a pattern recognition task \cite{velardo_automated_2015}. As we listen, we segment the music into motives, phrases, and sections. We identify similarities, repetitions, variations, and points of departure over time. Through this process, we form a mental representation of the music's structure. However, human memory's limitations become evident when individuals struggle to recall details; instead, they remember specific impactful moments \cite{ong_structural_nodate} and distinct fragments rather than entire compositions, highlighting the selective nature of musical memory \cite{miller_magical_1956,velardo_automated_2015}.

Given this understanding, songwriters strategically incorporate catchy repetitive phrases to enhance memorability and emotional impact for listeners \cite{guo_hierarchical_2021,livingstone_emotional_2012}. However, Burns \cite{burns_typology_1987} argues that repetition gains meaning through its relationship with change. Both endless repetition and constant variation could lead to monotony. Thus, repetition and change are opposite possibilities from moment to moment in music. The tension between them can be a source of meaning and emotion. For example, in pop music, the repetition of verse and chorus sections helps emphasize music ideas, while the contrast between verse and chorus can create more emotional intensity \cite{lu_meloform_2022}. Similarly,  within Western tonal music, artists leverage repetition of harmonic progressions (sequences of chords) to guide listeners through a journey that creates dramatic narratives, conveying a sense of conflict that demands a resolution \cite{lazzari_pitchclass2vec_2023,temperley_cognition_2004}.

While structure and predictability are key for many applications involving composition, cognitive science and musical therapy \cite{wigram_music_2006}, some contexts like video game and social media music require adaptive, event-driven approaches \cite{santos_adaptive_2022}. The music in video games often evolves in response to the player's progress, creating a unique journey for each player. In this scenario, long-term structural plans around musical form might be impractical, as the musical narrative needs to be flexible and responsive to the unfolding events in the game. Similarly, social media music quickly sets moods with catchy hooks rather than intricate structures \cite{frid_music_2020}. Additionally, avant-garde music is another area that deliberately challenges conventional musical structures, requiring artists to explore abstract and unpredictable forms that push listener expectations.

From a computational perspective, automatically analyzing and generating musical structure poses challenges for music research. Unlike other art forms like stories which have explicit plot structures, the “language” of music structure is more abstract and relies heavily on repetition, variation, and development of themes. This makes explicitly modelling musical structure difficult. While well-structured music generation with AI spans both raw audio and symbolic (sheet) music domains, this literature review specifically focuses on research pertaining to musical structures in the symbolic domain. While there has been a plethora of surveys on music generation systems, recent surveys such as \cite{briot2017deep,carnovalini2020computational,civit2022systematic,ji_survey_2023,liu2023literature,tang2022music} have covered works that use deep learning techniques. Among these, \cite{carnovalini2020computational} touches upon early deep learning methods built upon recurrent neural networks (RNNs), and generative adversarial networks (GANs), while placing greater emphasis on symbolic methods. Similarly, \cite{briot2017deep} also surveys early deep learning models until Music Transformer \cite{huang_music_2018}. Furthermore, \cite{liu2023literature} and \cite{tang2022music} have a very narrow focus on multi-track music using GANs and three specific deep learning models respectively. While \cite{ji_survey_2023} is comprehensive in their selection of papers by area of application, our literature review differs in two ways. First, we offer a deeper analysis on structure by exclusively focusing on studies that aim to improve musical structure in their methodology. Our selection of papers predominantly spans between 2007 to 2023. Early papers spanning this time frame can be seen in section \ref{symbolic-methods} which provides a foundational overview by covering symbolic methods before delving into later advancements using deep learning in section \ref{deep-learning-methods}. Second, we examine an emerging “subtask decomposition" approach in section \ref{sec:Subtask-Decomposition} that breaks down the music generative process into smaller steps incorporating neuro-symbolic methods or musical knowledge. Reflecting on this trend, we address future directions for modelling long term musical structure in section \ref{future-directions}.

\section{Symbolic Methods}
\label{symbolic-methods}
In early work on computerized music generation with long-term structure, \cite{cont_anticipatory_2007} proposes a reinforcement learning approach on top of a Variable-Length Markov model for musical style imitation and improvisation with long term dependencies. Multiple agents model different musical attributes (e.g. pitch, rhythm, harmony) to capture the anticipatory foundations of musical expectation in a collaborative yet competitive framework. The chosen “behavior” agent for each episode takes actions and updates the others’ policies accordingly. Agents leverage \textit{Factor Oracles}, which compactly gather repetitive sequence factors, enabling efficient access to long musical contexts. Results modeling Bach inventions demonstrates the model's sensitivity to phrase boundaries, which clearly segment the score into formal sections.

The authors of \cite{pachet2011markov} propose a constraint programming approach for controlled generation of Markov sequences, with applications to music. They formulate the Markovian property as a cost function represented by stacks of Elementary Markov Constraints and generate sequences by exploring the space of all sequences satisfying both the constraints and optimizing the Markovian cost. A key advantage of this approach is the ability to specify arbitrary structural constraints beyond just Markovian properties and find globally optimal solutions through search rather than greedy methods. These constraints can encode musical structures like repetition, motifs, endings, etc. Different Markovian cost functions can also bias generation towards more structured sequences. In addition, the model performs chunk-wise generation which allows it to maintain structure over long sequences (full songs) rather than just short clips. The chunks can be stitched to maintain longer-term dependencies. 

Another notable example of a Markov approach for generating music with long-term repetitive and phrasal structure is the Racchmaninof-Jun2015 algorithm \cite{collins_computer-generated_2017}. In this study, the authors develop a Markov model for inheriting long-term repetitive structure from template pieces using the SIACT pattern discovery algorithm from \cite{collins_comparative_2010}. The beat-relative-MIDI state space representation used in their paper is an improvement over a previous model called Racchmaninof-Oct2010 \cite{collins_developing_2016} that uses beat-spacing state space representation in which the spacing difference with respect to the previous note could consist of tonally obscure representations.  Using the extracted structural framework of a template piece, the Markov model can generate complete musical textures in a forward as well as backward generative way to form phrases that are perceived as having a beginning, middle and an end.

Herremans et al., \cite{herremans_generating_2015} present an application of Markov models combined with Variable Neighbourhood Search (VNS) to generate structured music in the style of Bagana, a traditional lyre from Ethiopia. The structure of Bagana pieces are encoded as cyclic patterns capturing repetitive cycles and global form. These patterns are realized by the VNS optimization technique that generates candidate melodies fitting the predefined structure of a template piece. A first order Markov model learned from the Bagana corpus is used in the next step to evaluate how well the melody matches the corpus statistics and musical structure. The VNS then selects the melodies optimizing the objectives to improve over iterations. While the Markov model does not directly generate notes, it provides the statistical basis for the objective functions guiding the VNS to produce melodies adhering to the structure with stylistic consistency. The integration of optimization and statistical learning helps in structuring music in Bagana's style.

Morpheus \cite{herremans_morpheus_2019} extends the VNS approach to polyphonic music, with automatic detection of more complex long term patterns in the template piece with respect to a user-specified tension profile. Morpheus comprises 3 components that are core to its framework. First, the authors use the COSIATEC and SIATECComPress \cite{meredith_cosiatec_2013} pattern detection algorithms to extract repeated note patterns from a template piece. Users can choose between COSIATEC, which captures each note in precisely one pattern, or SIATECCompress, which captures more relationships between different notes, resulting in overlapping patterns. This allows imposing long-term structure in the generated music. Second, these patterns are then constrained in a VNS optimization framework that assigns pitches to generate new music matching a target tension profile. And finally, tension used in the VNS optimization algorithm is quantified using the spiral array model of tonality developed in \cite{herremans_tension_2016}. While Morpheus has received favorable feedback overall, certain limitations are evident from a structural perspective. These include limited flexibility in pattern detection and the absence of captured variations or transformations, attributed to the pattern recognition algorithms used. Additionally, Morpheus is restricted in developing motifs, primarily repeating patterns without sophisticated development, and lacks a formal quantitative evaluation of structure retention, relying solely on informal listening tests for assessment. 

Genetic algorithms, a class of evolutionary computing approaches that are inspired by the principles of natural selection, have also been explored in the concept of evolving musical motifs and structures over generations. For example, in \cite{ting_novel_2017}, the authors propose an automatic music composition system called Phrase Imitation-based Evolutionary Composition (PIEC) that generates new melodies with structure and form by imitating phrases from a sample melody composed by people. The framework uses a genetic algorithm along with intraphrase and interphrase rearrangement to mimic the ascending or descending melodic progression and note distribution of phrases in the sample melody. Similarly, in \cite{alvarado_melody_2020}, the authors use a genetic algorithm to generate variations on a musical piece. To ensure that the generated variations have the underlying melodic skeleton of the original piece, the paper derives a melodic similarity fitness function from the spiral array model proposed in \cite{chew_real-time_2005}.

\section{Deep Learning Methods}
\label{deep-learning-methods}
Symbolic methods, though useful in their versatility, are limited in their development across diverse musical datasets because they rely heavily on specific templates and encode strict musical rules, potentially restricting the generated music. In contrast, deep learning breaks free from these constraints. By learning from extensive datasets, deep learning models understand complex patterns and functions, and exploit locality through the use inductive biases. This enables them to contribute to improved quality in applications such as prompt based conditional generation, unconditional generation with no inputs or assistance, musical infilling as well as accompaniment generation \cite{makris_conditional_2022}. 

\subsection{Foundational Deep Learning Techniques}
Lookback RNN and Attention RNN \cite{waite_e_generating_2016} by Google Magenta were two notable approaches early in the deep learning era that modeled long term structure in the generated melodies. Lookback RNN encoded repeating patterns from 1 or 2 bars ago in the input to allow the RNN model to identify repeating events better in the generated melody. Attention RNN used an earlier version of the self-attention mechanism from \cite{bahdanau_neural_2016} to generate melodies with longer dependencies without storing information in the LSTM cell's state. However, while these improvements are noticeable over brief periods, they may not necessarily impact the higher-level structural aspects of the piece.

Building on the Lookback RNN model, the authors of StructureNet \cite{medeot_structurenet_2018} propose an enhancement at the generation stage of an RNN based melody generation model that induces repetition and structure in the generated melodies. StructureNet is trained on sequences of structural elements like repeats extracted from a dataset of melodies. When generating new melodies, it runs alongside the melody RNN model, biasing it towards notes that would form repeats similar to those seen in the training data. To evaluate if StructureNet improves structure, the authors compared statistics related to repeats as well as pitch and rhythm distributions between melodies generated with and without StructureNet. The results showed that StructureNet increased the occurrence and lengths of repeats, matching the melodies in the training data more closely, while preserving pitch and rhythm characteristics.

In another study, \cite{chen_effect_2019}, the authors proposed using WaveNet \cite{oord_wavenet_2016}, a generative model based on dilated convolutional neural networks, for generating new melodies that fit a given chord progression. They found WaveNet was able to better learn musical structure compared to its LSTM counterpart. This is because the WaveNet model encodes melodic structure more explicitly through its dilated convolutions across multiple time scales. Each layer doubles the receptive field of the previous layer, allowing it to learn dependencies over larger spans of melody. In contrast, an LSTM relies solely on its internal memory to capture structure, which is less explicit. As a result, WaveNet was able to generate coherent rhythms and melodies over many time steps that exhibited more repetitive patterns and long-range dependencies over the LSTM baseline that was reflective of compositional structure. 

\subsection{Transformative Deep Learning Approaches}
Along the lines of model based enhancements for inducing structure in longer compositions is Music Transformer \cite{huang_music_2018}, which uses the transformer model equipped with the relative attention mechanism. This allows the model to learn patterns based on relative distances between events, making it easier to capture motifs and repetition on multiple timescales. However, this approach doesn't explicitly capture abstract structural elements such as how musical phrases and sections transition and develop over time. Additionally, when generating longer music pieces (over a minute), the model is not able to adhere to the initial prompt sequence and produces output with poor musicality and global structure.

In \cite{de_berardinis_modelling_2020}, the authors introduced Long-Short Term Universal Transformer (LSTUT), a fusion of transformers and RNNs, aiming to capture both local patterns and extensive musical relationships. This approach stemmed from a fundamental hypothesis: the intertwining of recurrence for short-term structure and attention for long-term dependencies enhances inductive biases, enabling the modelling of music spanning several minutes. However, while their methodology is promising, the evaluation is somewhat limited. Their assessment primarily relies on cross-entropy loss and manual inspection of attention heads to decipher the learned musical features. The lack of a subjective test and absence of generated samples restricts a comprehensive understanding of the model's potential. 

Along the lines of feature representation based enhancements is the Pop music transformer \cite{huang_pop_2020} in which the authors propose REMI (revamped MIDI derived events) to represent MIDI data following the way humans read them. REMI uses position and bar events to embed an explicit beat/bar grid, tempo events to allow flexible tempo changes, and chord events to represent harmony. This equips models with a sense of rhythm and harmonic structure to better learn musical dependencies. Pop Music Transformer uses REMI with a Transformer-XL backbone. Objective evaluation shows the model generates piano compositions with more consistent rhythm and clearer downbeats compared to baselines that use the MIDI-like representation \cite{oore2020time}. Subjective listening tests also indicate REMI results in more pleasing and coherent continuations of given musical prompts. While REMI offers a simple yet effective way to inject musical inductive biases into sequence models through data representation, the input sequence length can be a point of concern for music of longer duration.

The Theme Transformer model \cite{shih_theme_2022} effectively tackles the constraints of Music Transformer by incorporating thematic materials into the generated outputs. Initially, the authors employ an unsupervised approach, combining contrastive learning with clustering techniques to extract recurring thematic patterns from 2-bar melodic segments. These themes are then integrated into a transformer encoder, and music generation occurs through a transformer decoder, which cross-attends to the encoder. The architecture of the Theme Transformer incorporates gated parallel attention and theme-aligned positional encoding. This design ensures the sustained influence of the thematic condition over extended generation periods. This is achieved by modulating between the self-attention mechanism based on the decoder's auto-regressive inputs and the cross-attention from the encoder's thematic outputs. Evaluations demonstrate that the Theme Transformer excels in generating music with repeated thematic elements compared to prompt-based Transformers. However, it faces challenges in creating variations and evolving the recurring themes, showcasing areas for potential improvement.

In \cite{hu_beauty_2022}, a novel R-Transformer model is introduced, aiming to tackle the under-explored problem of modelling varied motif repetitions. To facilitate this, a new music repetition dataset comprising over 500,000 labeled motif repetitions across 5 types is constructed from an existing piano music dataset. R-Transformer combines a Transformer encoder for note representation learning and a repetition-aware learner that exploits repetition characteristics based on music theory, enabling controlled generation of designated repetitions. The model is trained using both reconstruction and classification losses. Evaluations show it outperforms baseline models in generating varied high-quality repetitions of motifs and creates subjectively more enjoyable music than prior models, as rated by both musicians and non-musicians. This is attributed to explicitly modelling repetition types and their combinations, which prior generative models overlook. 

\subsection{Hierarchical Neural Networks}
\label{sec:Hierarchical-Neural-Networks}
There exists a class of models that draw inspiration from the hierarchical nature of musical elements like notes, chords, bars, phrases, and movements. These models leverage this inherent hierarchy by employing specialized network structures to dissect and generate intricate melodic sequences. 

MeloNet \cite{hornel_melonet_1997} is an early example of a small feed-forward based neural network architecture that uses a hierarchical approach to generate melodic variations in Baroque style given a melody input. Specifically, it employs two neural networks operating at different timescales - a “supernet" that predicts motif classes at a higher level based on the melody and previous motifs, and a “subnet" that generates the actual notes based on the motif, harmony, and previous notes. 

Similarly, \cite{wu_hierarchical_2018} introduces a hierarchical RNN model for monophonic melody generation, employing three Long-Short-Term-Memory (LSTM) sub-networks that operate in a sequential manner: generating bar profiles, beat profiles, and notes. The outputs of higher-level sub-networks guide the generation of finer melody components in lower-level sub-networks. The subjective experiments conducted using this hierarchical approach demonstrate improved melodic quality and long term structure over MidiNet \cite{yang_midinet_2017} and MusicVAE \cite{roberts_hierarchical_2019}. 

Expanding on this concept, Guo et al. \cite{guo_hierarchical_2021} employed a three-tiered hierarchical RNN to simultaneously model rhythmic and pitch structures. This approach allowed the model to capture hierarchical musical relationships across varying time scales, evident in the presence of more repeated patterns indicated by a higher compression ratio metric. MusicFrameworks \cite{dai_controllable_2021} adopted a multi-step hierarchical process, breaking down melody generation into manageable subtasks guided by music frameworks. This method facilitated the use of simpler models trained with less data, enabling the creation of long-term structures, including repetitions, within full-length melodies. 

Another notable example of a hierarchical transformer model is HAT (\textbf{H}armony-\textbf{A}ware Hierarchical Music \textbf{T}ransformer) proposed in \cite{zhang_structure-enhanced_2022}. HAT leverages musical harmony to jointly model form and texture structure for pop music generation. HAT represents music as tokens with attributes like notes, chords and phrases, and uses Transformer blocks in a three tiered hierarchy to capture structure at different levels. Subjective as well as objective experiments on musical structural attributes show HAT generates pop music with improved coherence in form and stability in texture compared to prior methods. However, there is still room for performance improvement for longer chord progressions and ending sections.

\subsection{Long Sequence / Efficiency Models}
A class of models hypothesize generating longer sequences is key to improving musical quality and structure, a notion grounded in the observation that training data often surpasses the capabilities of standard full-attention vanilla transformer models. The Compound Word Transformer work \cite{hsiao_compound_2021} groups consecutive music tokens into “compound words" and processes them together for compactness. This allows their model to handle longer sequences to capture long term dependencies and induce structure. In a user study, the authors demonstrated improved structural outcomes during unconditional generation. Similarly, MuseNet \cite{christine_musenet_2019} adopts the Sparse Transformer \cite{child_generating_2019} architecture, expanding the context window to 4096 tokens. The aim is to generate music with extended structural coherence from a given prompt. Another notable advancement is Museformer \cite{yu_museformer_2022}, a Transformer model for long sequence symbolic music generation that captures musical structure through a fine and coarse-grained attention mechanism. It allows each token to directly attend to selected “structure-related" bars based on statistical analysis of common repetitive patterns in music, providing fine-grained attention for modelling local structures. All other non-structure-related bars are summarized via coarse-grained attention to retain necessary context while greatly reducing computation compared to full attention. This combined approach efficiently generates long, structured musical sequences by capturing repetitive patterns and variations through fine-grained attention, while still incorporating global context. However, one limitation is that the bar selection is based on statistics and may not generalize perfectly to all music. The SymphonyNet \cite{liu_symphony_2022} approach introduces a permutation invariant byte pair encoding (BPE) representation for symbolic symphony music generation. The BPE representation is proposed to preserve note structure through a compressed encoding scheme by preventing the overflow of long sequence multi-instrument tokens.

\subsection{Pre-training Methods}
Pre-training involves training a model on a large unlabeled dataset before fine-tuning it for a specific task, aiding the model in learning general representations applicable to downstream tasks. Pre-training can prove advantageous in symbolic music generation tasks due to the intricate, multi-dimensional structure of music, encompassing local and global patterns involving elements like pitch, rhythm and dynamics. Music's high complexity makes it important to learn musical syntax and patterns from vast datasets, enhancing coherence during fine-tuning. However, compared to other domains like text, there are fewer large-scale structured symbolic music datasets available \cite{makris_conditional_2022,sheng_songmass_2020,wu_melodyglm_2023} for task specific problems. Utilizing unlabeled data through pre-training facilitates transfer learning \cite{hosna2022transfer}, enabling models to leverage discovered patterns and structures when fine-tuning on limited task-specific datasets, maximizing the utility of available labeled data. A few studies \cite{donahue_lakhnes_2019,li_mrbert_2023,sheng_songmass_2020,wang_musebert_2021,wu_melodyglm_2023} have explored pre-training in the symbolic music domain aimed at generation with better musical structure and overall coherence.

In the LakhNES project \cite{donahue_lakhnes_2019}, the authors pre-trained a Transformer-XL model using language modelling on a large MIDI dataset. This was then fine-tuned on multi-instrumental music generation. The goal was to leverage the patterns learned during pre-training to generate more coherent long-form music. SongMASS \cite{sheng_songmass_2020} also used a masked language modelling approach to pre-train an encoder-decoder model. They specifically lengthened the masked spans during pre-training to capture longer-term repetitive structures like verses and choruses. This helped generate songs with improved global structure. MuseBERT \cite{wang_musebert_2021} and MRBERT \cite{li_mrbert_2023} adapted BERT \cite{devlin2018bert}, a language representation model for various downstream music generation tasks by pre-training on piano music segments. They proposed techniques to handle music's polyphonic and multi-dimensional nature compared to text. Recently proposed MelodyGLM \cite{wu_melodyglm_2023} introduces a multi-task pre-training framework to model both local patterns and long-term structure in melodies. It uses melodic n-grams and long span masking objectives tailored to music during pre-training. MelodyGLM also constructs a large-scale melody dataset to enable robust pre-training. Both objective and subjective evaluations demonstrate MelodyGLM's ability to generate melodies with improved structure and coherence.

\section{Subtask Decomposition}
\label{sec:Subtask-Decomposition}
In the context of deep learning, a new method has emerged which involves decomposing the training and generation process into smaller, more manageable steps (typically 2 stages). First, a high-level structure or plan for the music piece is created, outlining its organization and patterns. In the second stage, the actual musical content, such as notes and chords, is generated while adhering to this plan. This separation helps in addressing the challenges of modelling long-term structures and generating details locally. The plan ensures overall coherence, while the detailed music is filled in from the bottom-up, providing a holistic musical composition. There are two notable differences between pre-training and subtask decomposition methods. The former involves the use of the first pre-trained model in the fine-tuning stage whereas the latter involves an entirely separate model in the final stage. Second, the subtask decomposition framework explicitly models musical structure through repeated patterns or an outline of a melodic skeleton in its initial stage. Additionally, while subtasks could be thought of as a hierarchy, what separates them from the hierarchical models in section \ref{sec:Hierarchical-Neural-Networks} is that the latter trains a single model in a hierarchical fashion. In contrast, subtask decomposition approaches may involve multiple training objectives.

An early design of this framework can be seen in \cite{velardo_automated_2015} in which the authors use modular planning theory to generate tree like structures at multiple levels which are referred to as sentences, double phrases and phrases. Each level affects the next through object dependencies and constraints. These planning structures are then filled with musical material using a bottom-up Markov based generation technique. Their results show that automated planning can efficiently produce repetition and novelty in melodies, resembling human compositions.

In another study, \cite{wei_generating_2019} the authors focus on the problem of drum pattern generation, specifically drum sequences that are rhythmically and structurally compatible with a given melodic track. The proposed model uses a two stage generation framework. In the first stage, a variational autoencoder (VAE) generative adversarial network from \cite{larsen_autoencoding_2016} is trained to generate a self-similarity matrix (SSM) for the drum track given the SSM of the melodic track. This drum SSM captures the structural information of the music. In the second stage, another VAE model then generates the actual MIDI drum patterns conditioned on the drum SSM and using a bar selection mechanism to encourage self-repetition. Objective and subjective tests conducted against baselines yielded promising outcomes. Objective and subjective tests against baselines showed promising results, highlighting the model's enhanced performance due to its incorporation of global song structure through the drum SSM.

Similarly, PopMNet \cite{wu_popmnet_2020} proposes a two stage process for the generation of pop melodies with well defined long term structure. It represents melody structure as a directed acyclic graph capturing repetitive and sequential relationships between bars. In the first stage, a convolutional GAN model is trained on this graph based structural representation to generate plausible melodic structures. In the second stage, an RNN model is then trained to generate the actual melody conditioned on the structure and chords from the first stage. Human evaluations reveal that melodies generated by PopMNet receive higher ratings in terms of enjoyability, humanness, smoothness, and structure compared to previous models such as AttentionRNN \cite{waite_e_generating_2016}, LookbackRNN \cite{waite_e_generating_2016}, MidiNet \cite{yang_midinet_2017}, and Music Transformer \cite{huang_music_2018}. The analysis shows that PopMNet melodies exhibit repetition patterns similar to real pop songs, a feature previous models struggled to capture. However, one limitation of the model is its inability to capture complex structural patterns beyond repetitions and sequences.

Melons \cite{zou_melons_2021} overcomes PopMNet's limitations by utilizing a more complex multi-edge graph representation of musical structure, incorporating 8 types of bar-level relations compared to PopMNet's focus on repetition and rhythm sequences. Additionally, Melons adopts a Transformer architecture for both structure and melody generation, an upgrade from PopMNet's CNN and RNN models. In cases where no edges exist between bars, an unconditional transformer model generates melodies. Conversely, when relationships exist, a conditional transformer model is employed, considering the previous 8 bars and pairwise bar relationships as context. While Melons outperforms PopMNet in structure and overall musicality, there's still a gap between generated melodies and human-created music. About 80\% of the bars in the dataset used in the study cover the proposed relations, suggesting potential room for improvement by exploring additional structural relationships between bars.

In \cite{dai_personalized_2021}, the authors present a modular approach for generating stylistic pop music by imitating a single seed song. This approach uses purely statistical and rule-based methods to capture distinctive melodic, harmonic, rhythmic, and structural attributes from the seed song. A proposed structure alignment procedure helps adhere to the stylistic section lengths and repetition patterns of the seed song. Separate generative models produce melody lines, chord progressions, bass-lines, and overall song structure. The generated songs are evaluated using both objective similarity metrics and subjective listening tests. While results show the model can produce enjoyable, novel music that listeners recognize as similar to the seed song, the authors acknowledge the potential for enhancing musical quality by combining their knowledge-based method with advanced deep learning techniques. They note that much of the model's success stems from making good sequential choices guided by music theory and statistics. However, more sophisticated deep learning sequence models may be able to make even better choices for the next note in a generated sequence. The challenge would be retaining the benefits of the knowledge-based structure and harmony modelling while leveraging the pattern learning strengths of deep networks.

WuYun \cite{zhang_wuyun_2023} proposes a hierarchical two-stage skeleton-guided melody generation architecture that incorporates musical knowledge. It first extracts a melodic skeleton of structurally important notes using music theory concepts such as downbeats, accents and tonal tension. In the first stage, a Transformer-XL network is trained on the extracted melodic skeleton to generate new skeleton sequences. The second stage comprises a Transformer encoder-decoder model that generates the full melody conditioned on encoding the skeleton to guide the process. Experiments on the Wikifonia dataset and subsequent subjective evaluations show WuYun produces melodies with improved structure and musicality compared to prior models including Pop Music Transformer \cite{huang_pop_2020} and Melons \cite{zou_melons_2021} but still falls short with respect to the musical qualities of human compositions.

AccoMontage \cite{zhao_accomontage_2021} is a two step neuro-symbolic system for generating piano accompaniments for complete songs given a lead sheet that comprises the melody and chord information of the complete song. AccoMontage combines rule-based optimization for high-level structure and deep learning for local coherence. The system first retrieves candidate accompaniment phrases from a database using dynamic programming to optimize for good transitions and phrase-level fit. It then re-harmonizes the retrieved phrases using neural style transfer to match the target chord progression of the template song. Subjective and objective evaluations show AccoMontage generates more coherent and well-structured accompaniments compared to pure learning-based and template-matching baselines. However, the dependence on manual phrase level annotations in the dataset places constraints in the applicability of the model to other datasets. 

The TeleMelody model in \cite{ju_telemelody_2022} applies a two-stage neuro-symbolic approach for lyric-to-melody generation. The first stage trains a lyric-to-template model to generate the symbolic template from lyrics. The template design incorporates musical knowledge and rules revolving around musical elements such as tonality, chord progression, rhythm pattern, and cadence, providing a symbolic representation that bridges lyrics and melodies. The second stage then trains a template-to-melody model in a self-supervised manner by extracting these templates from existing melodies and training a transformer model to reconstruct the melodies from the templates. Evaluations show TeleMelody generates melodies with better structure and musical attributes than previous end-to-end models with less requirement on paired training data. This is possible due to the intermediate template which helps reduce overall difficulty and improve data efficiency.

MeloForm \cite{lu_meloform_2022} also uses a neuro-symbolic approach to generate melodies with musical form control by combining expert systems and neural networks. MeloForm has two main components. First, an expert system generates synthetic melodies with a given musical form using handcrafted rules. It develops motifs into phrases and arranges phrases into sections with repetitions/variations based on the musical form. Second, a 4 layered transformer encoder-decoder network refines the synthetic melodies from the expert system to improve musical richness without changing the form. The Transformer is trained on the Lakh MIDI Dataset \cite{raffel_learning-based_2016} using an iterative masked sequence-to-sequence approach. During refinement, it conditions on the adjacent phrase's rhythm and harmony to predict the pitch of masked phrases while preserving overall structure. The expert system provides flexibility over musical form, while the neural network improves musical coherence. This allows MeloForm to support various musical forms such as verse-chorus, rondo, variational, sonata, etc. Subjective evaluations performed on MeloForm's generated outputs show an improvement over Melons \cite{zou_melons_2021} in musical structure, thematic material, melodic richness and overall quality.

Compose \& Embellish \cite{wu_compose_2023} introduces another two-stage linear transformer based framework to generate piano performances with lead sheets as the intermediate output. The authors extract the monophonic melodic line using the skyline algorithm from \cite{uitdenbogerd_manipulation_1998}. For structural information, the authors employ a structural analysis algorithm from \cite{dai_automatic_2020} that utilizes edit similarity and the A* search to identify repetitive phrases within the composition. Notably, the first stage involves a pre-training strategy using a separate larger dataset (Lakh MIDI Dataset \cite{raffel_learning-based_2016}) for learning new lead sheets, setting it apart from the other two stage models. The second stage fine-tunes the lead sheet model and trains the performance model using the Pop1K7 dataset \cite{hsiao_compound_2021}. Conditioning the performance model on interleaved one-bar segments from the pre-trained lead sheet model enables the generation of complete performance bars with contextual understanding from the latest lead sheet bar. 

\section{Future Directions}
\label{future-directions}

\textbf{Exploring Structural Encoding}: While many music generation systems have used bar level representations such as REMI \cite{huang_pop_2020,wu_compose_2023} and REMI+ \cite{von_rutte_figaro_2022} for modelling music with structure, incorporating phrase level representations or meta data may be an interesting direction to explore. Recent work by \cite{naruse2022pop} extends the bar-level REMI encoding to incorporate phrase-level and bar countdown features. Their model demonstrates the value of encoding pop musical structure at the phrase level, aligning more closely with human music perception. However, phrase-based modelling and repetitive motif detection remains an open challenge in algorithmic music generation and composition. One barrier is the lack of standard practices or implementations for extracting musical phrases and structural annotations across genres. Moving forward, the development of robust tools to systematically extract and evaluate musical phrases across corpora would provide the necessary components to assemble hierarchical phrase structures, thereby enabling systems to construct forms at the phrase level. Additional annotations demarcating phrase boundaries within datasets would also facilitate progress in this area.

\textbf{Mastering Advanced Compositional Techniques}: Developing sophisticated long-term structure requires moving beyond basic repetitions and successions. Current models such as \cite{huang_music_2018} and \cite{shih_theme_2022} struggle with nuanced variations, hindering the conveyance of complex musical narratives. Advancements demand models to master musical development through compositional techniques such as fragmentation, inversion, augmentation, diminution, stretto, sequence, and modulation across various time frames. Learning how to manifest a motif across the piece in a more structural way by incorporating such melodic and rhythmic transformations may warrant for a similarity conscious generative framework in which the generator understands how similar the last generated phrase is to the initial motif, previous phrase and overall form and has the ability to go back to refine it. Building a pause-think-refine framework along similar lines to \cite{herremans_morpheus_2019} would help research advance beyond sequential note-to-note generations that do not explicitly model higher-level motivic structures and transformations.

\textbf{Integrated Neuro-Symbolic Approaches in Music Generation}: Music generation with long term structure is a complex endeavor that requires the coordination of multiple components at the preprocessing stage. For example, preprocessing data to extract melodic/rhythmic/phrase level features into meaningful musical representations provides the groundwork for modelling music, as seen in \cite{lu_meloform_2022,shih_theme_2022,wu_compose_2023,zhang_wuyun_2023}. However, recent advances in music generation have not yet effectively integrated insights from symbolic methodologies and cognitive musicology to achieve long-term structural coherence. While musical information retrieval has actively researched structural analysis algorithms for decades - spanning tempo estimation, key and modulation detection, cadence identification, and phrase similarity quantification, etc. - modern neural approaches surprisingly overlook these methods, opting for simplistic data preprocessing. Furthermore, cognitive studies elucidating how humans internalize and anticipate musical structure are imperative for designing evaluation protocols. Yet current assessments of generative systems' musicality, both objective and subjective, lack foundations in perceptual research. Effectively melding the methodological tools of symbolic analysis and lessons from music cognition remains an open challenge in paving the way for a truly unified neuro-symbolic approach. 

\textbf{Top Down Structure}: Sub-task decomposition studies in section \ref{sec:Subtask-Decomposition} such as \cite{wu_compose_2023,zhang_wuyun_2023,zou_melons_2021} clearly demonstrate the viability of top-down hierarchical plan-then-generate frameworks for algorithmic composition. By separating high-level musical form construction from lower-level sequence generation, these two-stage systems aim to improve global coherence. While we expect this trend to continue, we also hope to see studies that explore alternatives to the existing task breakdown involving pre-training, multi-task learning, unsupervised and contrastive learning methods. What data, features, representations, learning strategies and model architectures the sub-tasks consist of and how many stages of hierarchy make the framework most effective are  questions worth looking into. 

\section{Conclusion}
This literature review has charted elements of the evolution of modelling musical structure in AI-powered symbolic composition, from early symbolic systems to contemporary deep learning techniques and subtask decomposition frameworks. While progress is evident in capturing motifs, progressions, and global forms, modelling nuanced development and variation of themes much like human compositions across extended periods remains challenging. Key future directions involve moving beyond rigid structural units, integrating neuro-symbolic approaches at various stages in the system, and further exploring top down structures to shape musical narratives. Continued progress across complementary AI techniques provides hope for achieving the creative development of themes reflective of human composers and musical culture, which may someday be attainable.
 
\section*{Acknowledgements}
This work was supported by the UKRI and EPSRC under grant EP/S022694/1. We gratefully acknowledge the use of generative AI in drafting and revising certain sections of this literature review. Finally, we owe much appreciation to our reviewers for their insightful critiques which greatly strengthened the work.

\newpage

%
%
%
\bibliographystyle{splncs04}
\bibliography{references}

\end{document}